\documentclass[twocolumn,showpacs,amsmath,amssymb,floatfix]{revtex4-1}

\usepackage{graphicx}
\usepackage{dcolumn}
\usepackage{bm}

\makeatletter

\def\G4p{$\Gamma_4^{(2),+}$}

\makeatother

\begin{document}

\preprint{PrOs4Sb12-0.1}

\title{The Fermi volume as a probe of hidden order}

\author{A.\ McCollam$^{1,2}$}
\author{B.\ Andraka$^3$}
\author{S.R.\ Julian$^{1,4,*}$}
\affiliation{$^1$Department of Physics, University of Toronto, Toronto, ON, M5S 1A7, Canada.\\
$^2$High Field Magnet Laboratory, Institute for Molecules and Materials, Radboud University Nijmegen, 6525 ED Nijmegen, The Netherlands.\\ 
$^3$Department of Physics, University of Florida, P.O. Box 118440, Gainesville, Florida 32611-8440, USA.\\
$^4$Canadian Institute for Advanced Research,
    Quantum Materials Program,
    180 Dundas St.\ W., Suite 1400, Toronto, ON, M5G 1Z8, Canada.}

\date{\today}

\begin{abstract}
We demonstrate that the volume of the Fermi surface, 
  measured very precisely using de Haas-van Alphen oscillations, 
  can be used to probe 
  changes in the nature and occupancy of localized electronic states.
In systems with unconventional ordered states, this allows  
  an underlying electronic order parameter 
  to be followed to very low temperatures. 
We {describe this effect} in the field-induced antiferroquadrupolar (AFQ) 
  ordered phase of PrOs$_4$Sb$_{12}$, a heavy fermion intermetallic compound.  
We find that the phase of de Haas-van Alphen oscillations is 
  sensitively coupled, through the Fermi volume, to the configuration  
  of the Pr $f$-electron states that are responsible for AFQ order.  
In particular, the $\beta$-sheet of the Fermi surface expands or shrinks 
  as the occupancy of two competing localized Pr crystal field states changes.
Our results are in good agreement with 
  previous measurements, above 300 mK, of the AFQ order parameter by other methods. 
In addition, the low temperature sensitivity of our measurement technique 
  reveals a strong and previously
unrecognized 
  influence of hyperfine coupling on the order parameter below 300 mK within the 
  AFQ phase.  
Such hyperfine couplings could provide insight into the nature of 
  hidden order states in other systems. 
\end{abstract}

\maketitle

\section{Introduction}
\label{sec:intro}

Studies of correlated electron systems have led to
  the discovery of {many} novel ordered phases,
  sometimes characterized by subtle broken symmetries.
It is often challenging, however, to understand these ordered phases,  
  as it is 
  a common problem that no experimental probe couples in 
  a simple way to the microscopic order parameter. 
This makes it difficult to 
  determine how the order manifests itself in terms of 
  the quantum states of the electrons,  
  and sometimes even to determine how strong the order is. 
For example, the true nature of spin arrangements in 
  antiferromagnets was ``hidden" for several decades, until the development of 
  neutron diffraction as a probe that couples to the order parameter \cite{shull51}. 
There are several modern examples of hidden order, which are among the most 
  active topics of investigation in the field of strongly   
  correlated electron systems. 
These include electronic nematics -- states that are characterized macroscopically as 
  translationally invariant electronic states that break the rotational 
  symmetry of a crystal -- 
and exotic superconducting states.
 It is believed that nematic states arise from fluctuating 
  density waves, but in systems of interest such as 
  the bi-layer ruthenate Sr$_3$Ru$_2$O$_7$,  
  quantum Hall systems, and some cuprate and iron-pnictide superconductors,  
  the microscopic nature of these broken-symmetry electronic states has not 
  been directly observed \cite{fradkin10}.  
In 
systems such as 
the heavy fermion superconductor UPt$_3$, 
  the order parameter 
  can be difficult to determine, because there are no 
  available probes that couple directly to the gap function  \cite{joynt02}. 

Electric multipole order of the kind observed in PrOs$_4$Sb$_{12}$ is a type of 
  hidden order
%
widely observed in $f$-electron materials \cite{hanzawa83,yamauchi99,morin90}, 
  in which 
  the electron density around some atoms in the unit cell spontaneously distorts in a
  repeating pattern throughout the crystal.  
The change in electron density
  is a very small perturbation of the total electron density,
  so that traditional probes such as x-rays couple only
  very weakly. 
Neutron scattering may couple indirectly to multipolar order if
  there is an admixture of 
  magnetic dipole as well as charge order, or if there is a sufficiently 
  large lattice distortion \cite{Walker94}
  but there are cases of quadrupolar order where neutron diffraction patterns 
  are unchanged upon entry into an AFQ phase 
  \cite{yamauchi99}. 
A well-known example of a system with 
  hidden order that may be due to multipolar 
  charge ordering is URu$_2$Si$_2$, for which several different varieties 
  of multipolar order, from quadrupolar to hexadecapolar, as well as a number 
  of other scenarios, have been proposed \cite{mydosh11}. 
Here, we show that 
  precise measurements of the volume of the Fermi surface 
  can reflect the modified charge density of multipolar-ordered states,
  allowing such order to be measured at very low temperatures. 

\begin{figure}
\includegraphics[width=8.0cm]{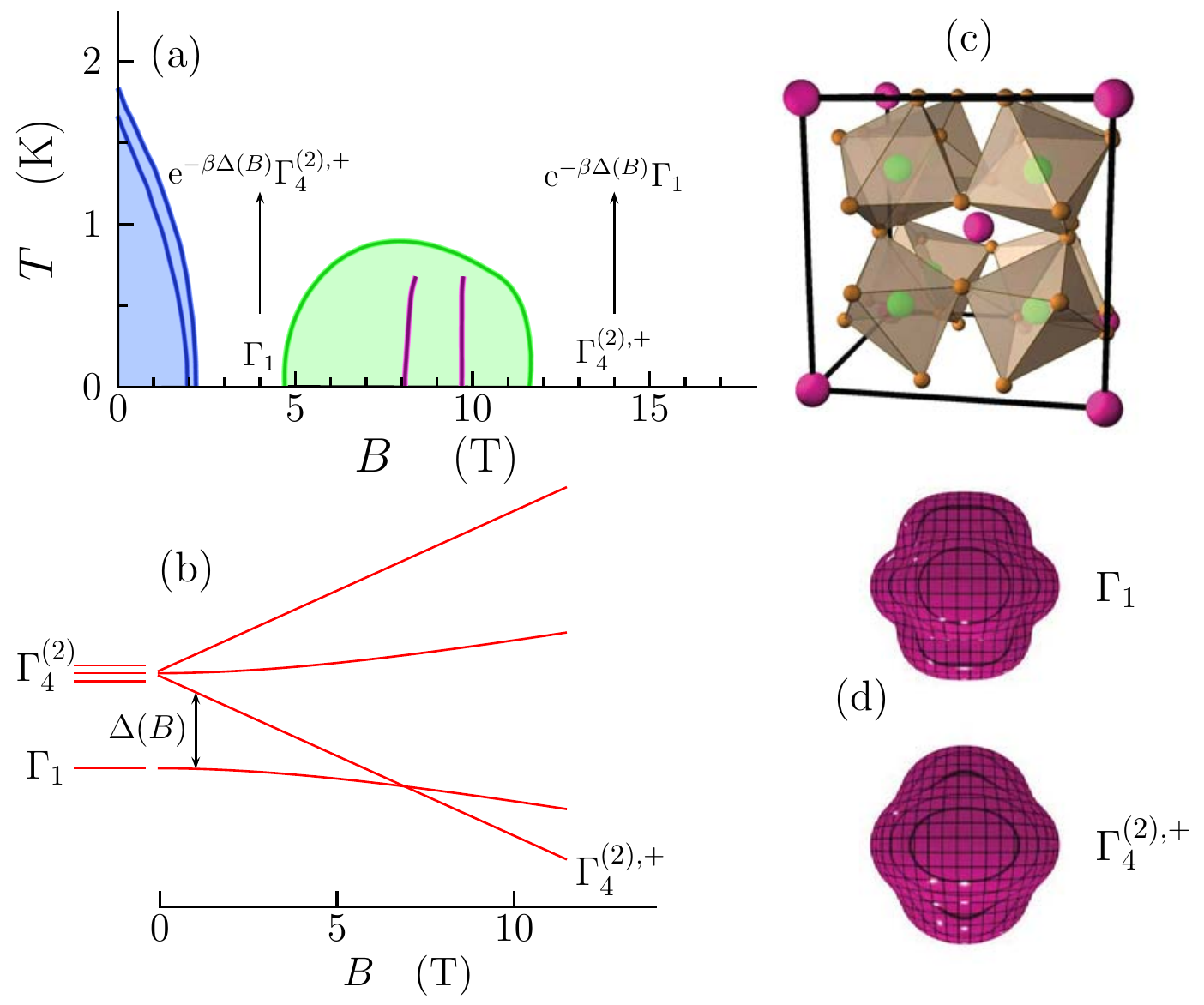}
\caption{ Antiferroquadrupolar order in PrOs$_4$Sb$_{12}$. 
(a) $B-T$ phase diagram of PrOs$_4$Sb$_{12}$, based on reference \cite{tayama03}.  
The blue region is superconducting, with a 
  double superconducting transition indicated by the blue lines.  
The green line denotes the AFQ phase boundary for $B \parallel (110)$, 
  and the two purple lines are transitions within the AFQ phase, 
  believed to involve changes in the 
  ordered structure \cite{tayama03}. 
Below 4.7 T, for $B \parallel(110)$, $\Gamma_1$ is the ground state, 
  with a field dependent gap $\Delta$ to the quadrupolar excitations 
  involving $\Gamma_4^{(2),+}$ 
  as indicated in (b).  
($\beta = 1/k_B T$). 
Above 11 T this situation is reversed.  
In the AFQ-ordered region, the ground state involves 
  coherent superpositions of $\Gamma_1$ and $\Gamma_4^{(2)}$ on each site. 
(c) shows the crystal structure of PrOs$_4$Sb$_{12}$, with Pr atoms in pink, 
  Os in green and Sb in orange.  
(d)  Charge distribution 
  of the Pr 4$f$ electrons in the two lowest-lying crystal field levels. 
The charge distributions are similar in shape, and 
  mostly buried within the ionic radius of Pr$^{3+}$, so they are only 
  weakly felt by the conduction electrons on the surrounding Os and Sb ions. 
 }
\label{fig-phasedia}
\end{figure}

In multipolar order, degenerate or nearly degenerate localized electronic 
  states form superpositions that lower the energy of the ground state. 
In PrOs$_4$Sb$_{12}$, the material we study,  
  antiferroquadrupolar (AFQ) order 
arises in the doubly occupied $4f$-electron shell of the Pr ions.
The AFQ phase appears for applied magnetic fields between about 4 and 12 T 
  and temperatures below $\sim$1 K \cite{aoki02} 
  (the green region in Fig.\ \ref{fig-phasedia}a)
%
and is understood as follows \cite{shiina04}.
In the skutterudite crystal structure of PrOs$_4$Sb$_{12}$, shown in 
  Fig.\ \ref{fig-phasedia}c, 
  the $J = 4$ spin-orbit-coupled ground state of a Pr 4$f^2$ state 
  feels the electric field of 
  the surrounding cage of 12 Sb ions, 
  lifting the 9-fold degeneracy it would have in free space. 
The resulting localized crystal field states have a ground 
  state singlet, labeled $\Gamma_1$, and a very low-lying triplet, 
  labeled $\Gamma_4^{(2)}$. 
(The admixtures of the various $|J,J_z\rangle$ states 
  in $\Gamma_1$ and $\Gamma_4^{(2)}$ are given in 
  references \cite{aoki02,kohgi03,shiina04,shiina04b}.) 
All other crystal-field levels are at much higher energies and can
  be ignored.
(For more details see Appendix 2.)

Application of a magnetic field causes the ground state to acquire a small 
  admixture of the $m_z = 0$ triplet state, $\Gamma_4^{(2),0}$, 
  and, more importantly, it splits the $m_z = \pm 1$ members of the triplet state, so that 
  the spin-up branch of the triplet, $\Gamma_4^{(2),+}$, crosses 
  $\Gamma_1$ near 8 T \cite{shiina04,aoki02}, as shown schematically in 
  Fig.\ \ref{fig-phasedia}b. 
The charge distributions for $\Gamma_1$ and $\Gamma_4^{(2),+}$ are pictured in 
  Fig.\ \ref{fig-phasedia}d. 
A weak nearest-neighbour interaction between the quadrupole moments of 
  these distributions,  
  combined with the near-degeneracy of
  $\Gamma_1$ and $\Gamma_4^{(2),+}$ close to 8 T, leads to the new AFQ 
  ground-state 
  in which there is an electric quadrupole moment that alternates on 
  neighbouring sublattices, due to a 
  coherent superposition of the $\Gamma_1$ and $\Gamma_4^{(2)}$ states on 
  each site. 
Within mean-field theory, this superposition would have the form 
  $\sum_{i,j}a_{i,j} |\Gamma_4^{(2),j}\rangle+b_i |\Gamma_1\rangle$, 
  where $i=1,2$ is a sublattice index and $j=+,0,-$ sums over the three states 
  in the triplet.
Outside the AFQ phase, in contrast, the $\Gamma_1$ and $\Gamma_4^{(2),+}$ states 
  would be randomly populated according to the Boltzmann distribution as indicated in 
  Fig.\ \ref{fig-phasedia}(a).
Both within and outside the AFQ phase, the occupancies of
  $\Gamma_1$ and $\Gamma_4^{(2),+}$ change rapidly with field and temperature.

The basic outline of this picture has been demonstrated with magnetization and 
  neutron scattering measurements, and extensive theoretical work
  \cite{shiina04,tayama03,kohgi03,shiina04}:
  $\Gamma_4^{(2),+}$ carries a magnetic moment and 
  $\Gamma_1$ does not, 
  thus the occupancy of the \G4p state can be determined by 
  magnetization measurements, while neutron scattering measurements find 
  that, in addition to the component of 
  magnetization parallel to the applied magnetic field $B$ 
  there is, within the AFQ phase, a comparatively weak antiferromagnetic 
  moment perpendicular to $B$ \cite{kohgi03,kaneko07}. 
In this paper, we show that changes in the population of 
  the two crystal-field states produce
  a small but clearly observable effect on the 
  size of one of the Fermi surfaces, 
  so that it expands or contracts as the relative occupation 
  of $\Gamma_1$ and $\Gamma_4^{(2),+}$ changes. 
This effect is measurable in the de Haas-van Alphen (dHvA) effect 
  via the Onsager formula relating the 
  dHvA frequency $F$ to the extremal Fermi surface area $\cal A$, 
  $F = \hbar {\cal A}/2\pi e$, 
  and means that 
  quantum
  oscillations can provide new insight into the temperature and field
  dependence of the quadrupolar states. 
The dHvA effect has played a prominent role in condensed matter physics  
  since it's discovery over 80 years ago  
  but this capability has not previously been exploited
  and may have 
  important applications, in particular because of the excellent 
  sensistivity of the dHvA technique at millikelvin temperatures.   

\section{Experiment} \label{sec:experiment}

Our measurements were carried out on a 
  single crystal of PrOs$_4$Sb$_{12}$ weighing 40 mg and having 
  dimensions $1.7\times 1.6\times 2.12$ mm$^3$.
The crystal was grown by a standard Sb-self-flux growth method.  
The residual resistivity
ratio, 
defined as the ratio of the zero-field resistance 
  at room temperature to that extrapolated to absolute zero,  
  measured on samples from the same batch, 
  fell in the range 70 to 80.
The crystal was heat-sunk to the mixing chamber 
  of a dilution refrigerator
  through an annealed silver wire that was 
  soldered to one corner of the sample.
The dHvA effect was measured using the standard field-modulation 
  technique with second-harmonic detection, with  
  the sample in an astatic pair of pick-up coils.
The modulation frequency was 6 Hz, with modulation field 
  amplitudes of 0.0126 T for fields from 2.5 to 8 T, and 0.021 T for 
  fields from 7 to 18 T. 
The signal from the pick-up coils 
  was measured using a lock-in amplifier, via  
  a low-temperature transformer with a turns 
  ratio of approximately 100, and a low noise preamplifier.
The sample and pick-up coils were 
  placed in a graphite rotation mechanism with a rotation range of 
  approximately 90$^\circ$.  

Measurements were performed with $B$ parallel to both the (110) and (100) 
  crystallographic directions. 
In this paper, only the former measurements are reported because 
  $T_{AFQ}$ is lower along the (110) direction, 
  and quantum oscillations were therefore better resolved across the
  entire AFQ phase for $B \parallel (110)$.
The results along (100) were 
fully
consistent with what we report here. 
Measurements were performed for magnetic fields between 2.5 and 18 T, 
  and at temperatures $T$ from 30 mK to 2.5 K, 
  a sufficiently wide range of
  magnetic fields and temperatures that the oscillations can be followed
  across the entire AFQ phase.

\section{Results}
\label{sec:results}

\begin{figure*}
\includegraphics[width=14.0cm]{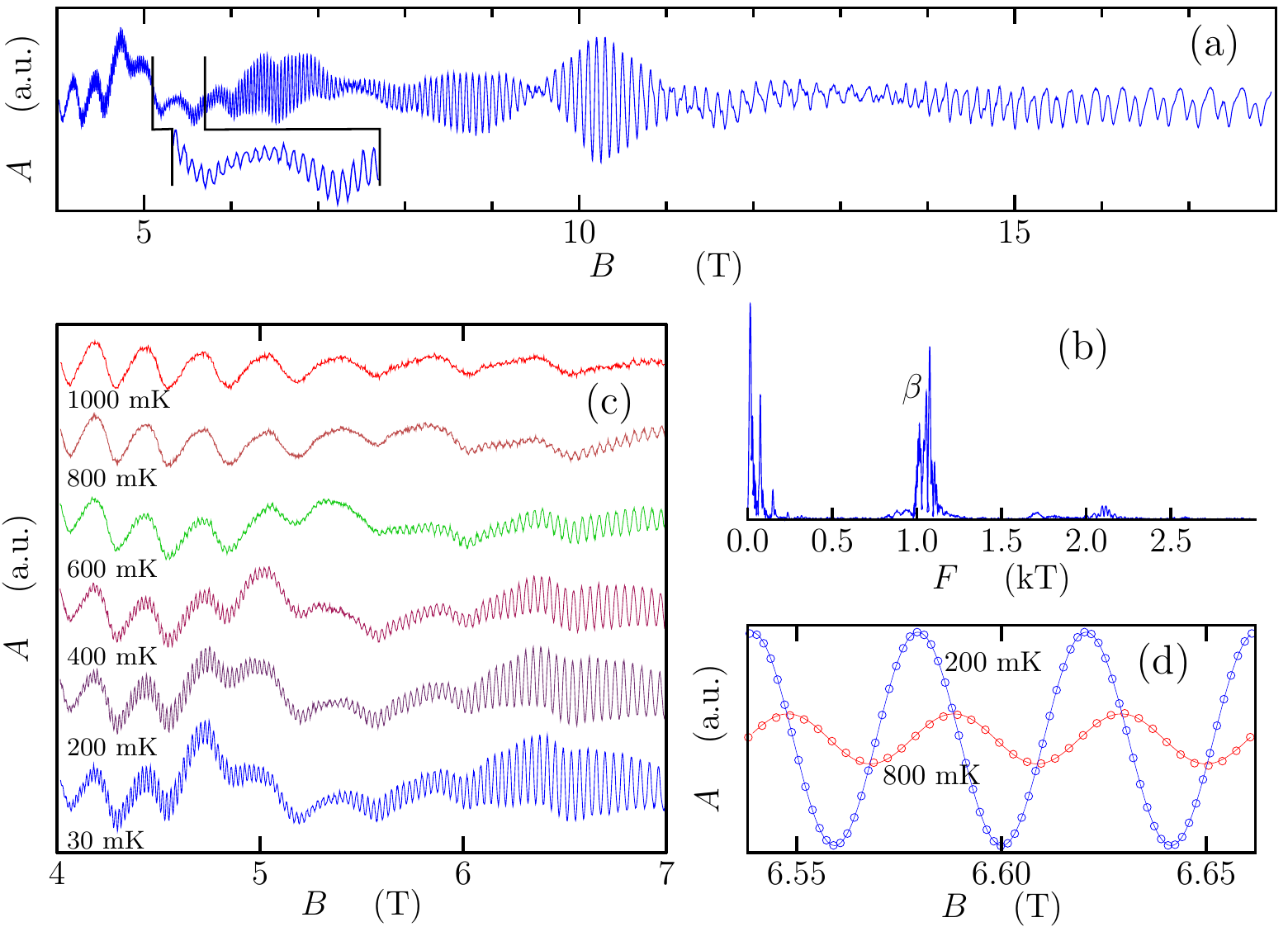}
\caption{
Typical quantum oscillations from PrOs$_4$Sb$_{12}$.  (a) shows 
data from 18 T 
  to 4 T.  
(b) shows the Fourier spectrum of the trace in (a), after the data have been 
  replotted as a function of $1/B$.  
  The $\beta$ frequency corresponds to the broad peak above 1 kT.  
(c) shows data from 4 to 7 T, between 30 mK and 1 K, to illustrate the temperature 
  dependence of the oscillation amplitudes. 
(d) shows the results of fitting 
Eq.~(\ref{eq:fit})
to narrow sections of the data at 6.6 T.  
  Note that there is a phase shift between the 200 mK and 800 mK oscillations.  
The data in (d) have been filtered so that only frequencies between 
  0.5 and 1.5 kT are present. 
  The solid lines are the fitted curves. 
 }
\label{fig-oscs}
\end{figure*}

Figure \ref{fig-oscs} shows typical results of our dHvA measurements 
  on PrOs$_4$Sb$_{12}$.  
Data at 100~mK, for 
magnetic
field between 18 and 4 T 
applied along
the (110) 
  crystallographic direction, are presented in Fig.\ \ref{fig-oscs}a. 
This demonstrates that we observe strong quantum 
  oscillations across the AFQ phase, which lies between 4.7 and 11.6 T. 
Figure \ref{fig-oscs}b shows the Fourier transform of the sweep in (a); 
  the broad structured peak labelled $\beta$ corresponds to the high-frequency oscillation 
  in Fig.\ \ref{fig-oscs}a, and 
  arises from a small, roughly spherical, 
  hole-type Fermi surface, first observed by Sugawara et al.\ \cite{sugawara02}. 
The electronic states of this Fermi surface arise predominantly from $p$-orbitals on 
  the cage of Sb ions that surrounds each Pr ion.  

dHvA oscillations arise from periodic (in inverse applied magnetic field) variations in the 
  orbital diamagnetism of conduction electrons as the density of states 
  changes due to quantized Landau levels passing through the Fermi energy 
\cite{shoenberg84}. 
In conventional metals the quantum oscillatory magnetization takes the form 
  \begin{eqnarray} 
  \tilde{M} = \sum_{p=1}^\infty A_p(T,B) \sin\left(2\pi p \frac{F}{B} + \phi\right), \label{eq:LK}
  \end{eqnarray}
  where $\phi$ is a constant phase, $F = \hbar{\cal A}/2\pi e$ with ${\cal A}$ being the 
  extremal cross-sectional 
  area of the Fermi surface, and $p$ is the so-called harmonic number. 
All of the measurements in this paper have focused on the $p=1$ term. 
The amplitude $A_p(T,B)$ is given by the Lifshitz-Kosevich expression 
  \begin{eqnarray}
    A_p(B,T) &=& {K}\frac{B^{5/2}}{p^{1/2} m^*}\left| 
        \frac{\partial^2 {\cal A}}{\partial k_z^2}\right|^{-1/2}  
        R_{p,\sigma}\, {\rm e}^{-\pi pr_c/{l}_\circ}\, \frac{X_p}{ {\rm sinh}X_p} \nonumber \\
  & & {\rm where} \qquad
        X_p = \frac{2\pi^2 p k_B T}{\hbar\omega_c}.  \label{eq:TLK}
  \end{eqnarray}
%
The important factors in this expression are: the spin damping term
 $R_{p,\sigma}$, which arises from the interference of the quantum 
oscillations from the spin-up and spin-down branches of the Fermi surface;
the Dingle factor ${\rm e}^{-\pi r_c/l_\circ}$, which accounts for 
damping of quasiparticles by scattering, where $r_c$ is the cyclotron radius 
and $l_\circ$ is the quasiparticle mean free path; and the cyclotron frequency
$\omega_c = eB/m^*$. $k_B$ is Boltzmann's constant. 
%
The $X/{\rm sinh}X$ 
term
allows the quasiparticle effective mass, 
  $m^*$, to be determined from the temperature dependence of quantum oscillations. 
We will discuss the effective masses in PrOs$_4$Sb$_{12}$ 
elsewhere,
but it should be noted that 
  this $X/{\rm sinh}X$ term falls to zero when $k_BT \gg \hbar\omega_c$. 
The fall in amplitude of the $\beta$ oscillation with increasing temperature 
  can be seen in Fig.\ \ref{fig-oscs}c. 
This thermal damping of the oscillations  
  restricts our observations to low temperature, 
  imposing a field-dependent temperature cut-off that rises from about 0.8~K at 
  4~T to about 3~K at 18~T. 

Normally, a Fourier spectrum such as Fig.  \ref{fig-oscs}b consists of sharp peaks
  arising from well-defined extremal areas of the Fermi surface, and, 
  indeed, sharp peaks corresponding to the low frequency oscillation of 
  Fig~\ref{fig-oscs}a can be seen below 0.3 kT. 
The $\beta$ peak, however, is a broad clump of peaks. 
This is not due to a superposition of many frequency components, 
%
but,  
as we 
shall
show, 
  is because the Fermi surface area is changing non-linearly, and by a large amount 
  as the magnetic field is swept.  
As a result, Fourier analysis is not a useful way of analyzing 
  the $\beta$ oscillations in PrOs$_4$Sb$_{12}$. 
Instead, we have adopted the simple procedure of 
  fitting the oscillations over narrow magnetic field ranges, just three periods  
  wide, with a function 
of the form
\begin{eqnarray} 
 A_f \sin\left( 2\pi F_f /B + \phi_f \right),     \label{eq:fit}
\end{eqnarray}
which is based on the Lifshitz-Kosevich expression (\ref{eq:LK}).
$A_f$, $F_f$ and $\phi_f$ are the fitted dHvA amplitude, frequency and phase.  
Fig.\ \ref{fig-oscs}d shows an example of such fits to sections of the data 
  centred on 6.6~T, at 200 and 800 mK. 

\begin{figure}
\includegraphics[width=8cm]{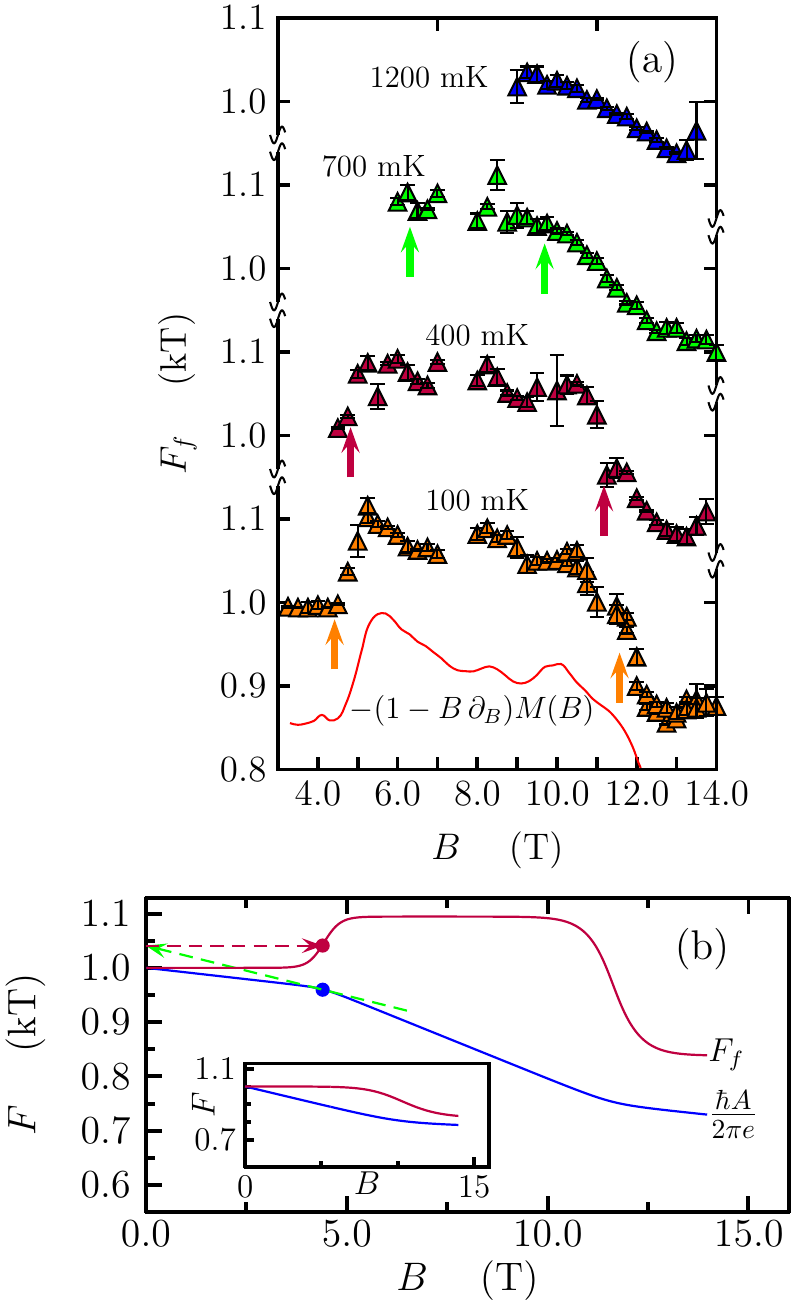}
\caption{ The magnetic field dependence of the dHvA frequency. 
(a) dHvA frequency $F_f$ vs.\ $B$ at temperatures from 100 mK to 1.2 K.
The values of $F_f$ are extracted from fits of Eq.\ \ref{eq:fit} 
  to sets of three periods, as described in the text. 
The arrows indicate the boundaries of the AFQ phase \cite{tayama03}.
The absence of low-field data at high temperature is due to the
  dHvA thermal damping factor in the Lifshitz-Kosevich amplitude, Eq.\  \ref{eq:TLK}.  
The red 
curve at the bottom is proportional to $-(1 - B \frac{\partial}{\partial B})M$, 
  where $M$ is the magnetization derived from the 
  60 mK $M$ vs.\ $B$ data of reference \cite{tayama03}. 
(b) Main figure: simple model curves showing a hypothetical low-temperature 
  field dependence of the Fermi 
  surface area ${\cal A}$ (blue line) together 
  with the back-projected frequency $F_f$ that would be seen in 
quantum oscillation
measurements (red line).  
The dashed green line shows how $F = \hbar{\cal A}/2\pi e$ at 4.4 T  
  is back-projected to produce the measured frequency $F_f$ (red dot). 
The inset shows a hypothetical field dependence of $\hbar{\cal A}/2\pi e$ and the 
  corresponding measured dHvA frequency $F_f$ at 
  a temperature above the maxiumum AFQ phase transition temperature.
The hypothetical field dependences of ${\cal A}$ were chosen to produce $F_f$ curves 
  that resemble the measured curves in (a) at 100 mK and 1200 mK. 
 }
\label{fig-FvsB}
\end{figure}

This fitting procedure allows us to extract the 
  field and temperature dependence of the dHvA frequency. 
Figure \ref{fig-FvsB}a shows the $B$ dependence of $F_f$ at 
  100, 400, 700 and 1200 mK.
In this figure, the arrows indicate the approximate boundaries of the
  AFQ phase according to reference \cite{tayama03}. 
(Note that 1200 mK is above the maximum AFQ phase transition
  temperature.)
In the 100~mK data 
of Fig.\ \ref{fig-FvsB}a 
(orange triangles) it can be seen that $F_f$ jumps 
  up upon entry into the AFQ phase, and jumps down upon exiting 
  the phase near 12 T, 
  establishing that the Fermi surface is sensitive to the AFQ order.

When an extremal area of the Fermi surface is 
  magnetic field dependent,  
the measured dHvA frequency 
cannot be interpreted
using the Onsager 
  formula, 
  $F = \hbar{\cal A}/2\pi e$ 
  (as explained
in Appendix 1). 
Instead, the measured frequency is the `back-projection' of $F$,
  given by $F_f = (1 - B\partial/\partial B)F$.
Geometrically, $F_f$ is the intercept at $B=0$ of the tangent 
  to $F$, illustrated in Fig.\ \ref{fig-FvsB}b for the point at 
  4.4 T (blue dot), which back-projects along the dashed green line, to give 
  the value of $F_f$ shown by the red dot on the upper curve. 

Thus, rather than the Fermi surface expanding and contracting at the 
  AFQ phase boundaries, as the 100 mK 
  data in Fig.\ \ref{fig-FvsB}a would imply if the Onsager formula 
  $F = \hbar{\cal A}/2\pi e$ were applied, 
  we believe that 
it
follows something like the 
  blue curve in the main panel of Fig.\ \ref{fig-FvsB}b: 
  the Fermi surface contracts monotonically with increasing field, but it does so 
  more rapidly within the AFQ phase, producing a 
  a back-projected frequency (red line) that 
  jumps up upon entering the AFQ phase.  

At the bottom of Fig. \ref{fig-FvsB}b, below the 100~mK data, is a red
curve showing the behaviour of  $-( 1 - B\partial/\partial B) M(B)$, 
where $M(B)$ is taken from published magnetisation measurements at 60~mK
\cite{tayama03}. 
At high field and low temperature $M(B)$ 
  measures the average occupancy of the \G4p crystal field state, 
  because \G4p has a magnetic moment while $\Gamma_1$ does not. 
The 
close
correspondence between $-(1 - B\partial/\partial B)M(B)$ at 60 mK and
  our $F_f(B)$ data at 100 mK strongly suggests that ${\cal A}(B)$ is
  proportional to $-M(B)$ (or more precisely, the {\em change}
  in ${\cal A}(B)$, ${\cal A}(B) - {\cal A}(0)$, is proportional to $-M(B)$).
This, in turn, tells us that ${\cal A}(B)$ is also dependent on the occupancy of 
  \G4p. 
Our main conclusion from Fig.\ \ref{fig-FvsB}a is, therefore, that the $\beta$ sheet of the 
  Fermi surface shrinks as the occupancy of \G4p grows relative to that of $\Gamma_1$.


\begin{figure}
\includegraphics[width=8cm]{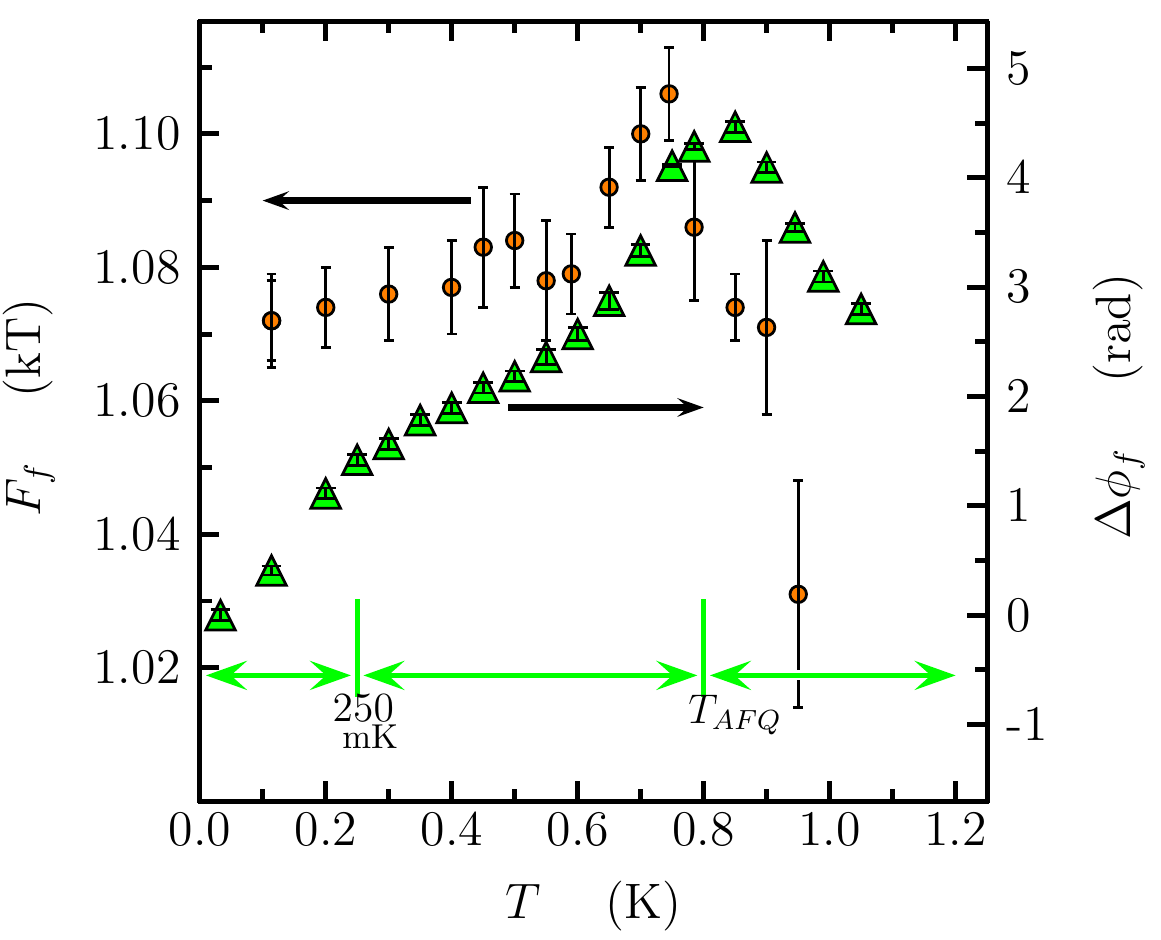}
\caption{ Temperature dependence at 6.2 T of the dHvA frequency $F_f$ (circles) and  
  the phase $\Delta\phi_f(T)$ (green triangles).  
Both show a peak near the antiferroquadrupolar phase transition, $T_{AFQ}$ (taken from 
  reference \cite{tayama03}), 
but the phase data are less noisy, and they reveal 
  an additional downturn below 
%
$\sim$ 250 mK that is not evident in the backprojected 
  frequency $F_f$. 
}
\label{fig:FvsT}
\end{figure}

\begin{figure}[h]
\includegraphics[width=6.6cm,height=15.0cm]{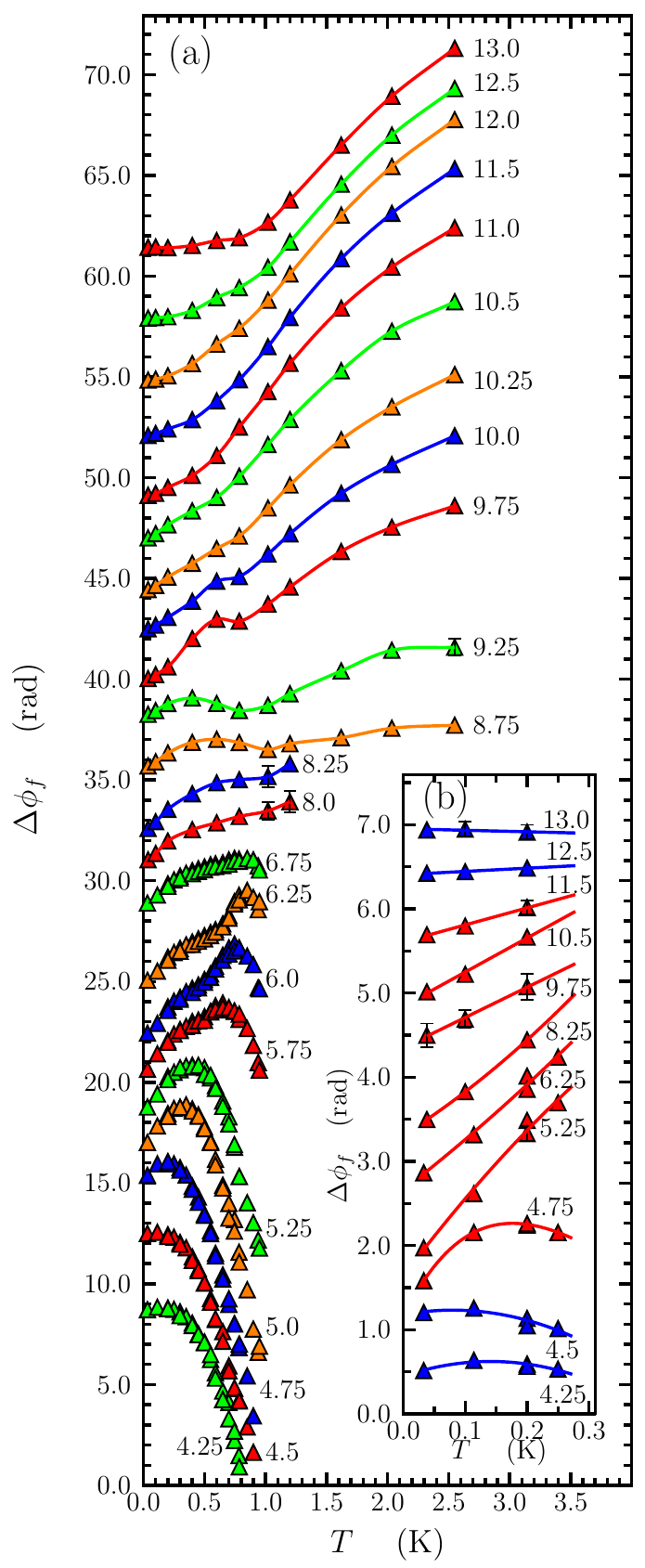}
\caption{The temperature dependence of $\Delta \phi_f(T)$ 
at magnetic fields spanning the AFQ phase.  
Because of the dHvA thermal damping factor (see Eq.\ \ref{eq:TLK}),
  below 8~T we only have data up to 950~mK.  
Error bars are statistical errors from averaging several sweeps with 
  identical conditions. Where errors are not plotted, the error bars 
  are smaller than the point size.
The lines are guides to the eye, and  
  the curves are shifted vertically for clarity; without this shift the 
  lowest temperature point would be at 0 radians for every curve.
The inset, (b), focuses on data below 300 mK for selected curves from (a). 
The blue(red) data are outside(inside) 
the AFQ phase. 
The lines are guides to the eye.
}
\label{fig-PhivsT}
\end{figure}

We turn now to the {\em temperature} dependence of the Fermi surface, which is 
  the primary focus of this paper.  
In Figure \ref{fig:FvsT}, the small circles 
  show the temperature dependence of $F_f$ at 6.2 T. 
At this field, the thermally driven AFQ phase transition, $T_{AFQ}$, is near 
  0.8 K. 
There is a peak in $F_f(T)$ 
  near this temperature, but the error bars are large, the data are rather noisy, 
  and it is not simple to connect the temperature dependence of the back-projected 
  frequency to the temperature dependence of the Fermi surface.  

The 
triangles in Fig.~\ref{fig:FvsT} illustrate another 
  approach to the data, in which we 
  consider the temperature dependence of the dHvA {\em phase}.
That the phase is temperature dependent can be seen in 
  Fig.\ \ref{fig-oscs}d: there is roughly a 
  $\pi/2$ phase shift between 200 mK and 800 mK. 
We extract this phase shift from the data by fitting
Eq.\ \ref{eq:fit} to three  periods 
  surrounding the field of interest 
(6.2 T for the data in Fig.~\ref{fig:FvsT})
at the base temperature of our measurement, 
$T_\circ$, 
using
$A_f$, $F_f$ and $\phi_F$ as free parameters.
This fit yields
  $A_f(T_\circ)$, $F_f(T_\circ)$ and $\phi_F(T_\circ)$. 
At all higher temperatures ($T > T_\circ$),
 $F_f$ is held fixed at its base temperature value, $F_f(T_\circ)$, 
  and only $A_f(T)$ and $\phi_f(T)$ are free parameters
in the fits.
The temperature dependent phase shift is then 
\begin{eqnarray}
\Delta\phi_f(T)  = \phi_f(T)- \phi_f(T_\circ). 
\end{eqnarray}

Fig.\ \ref{fig:FvsT} shows that 
  $\Delta\phi_f(T)$ is less noisy than $F_f(T)$, 
  with error bars that are smaller than the data points, and 
  with a clear peak near $T_{AFQ}$. 

The interpretation of $\Delta\phi_f(T)$ is straightforward:\cite{lonzarich74}
If the temperature-dependent change of $\cal A$ is 
  small compared to $\cal A$ itself 
  then, at a given field $B_\circ$,  
\begin{eqnarray}
\Delta\phi_f(T,B_\circ) 
             =  \frac{\hbar}{eB_\circ} \, \Delta {\cal A}(T,B_\circ).  \label{eq-deltaphi}
\end{eqnarray}
That is, the temperature dependent phase shift 
  directly gives the temperature dependent change of the Fermi surface extremal area.
This is shown in Appendix 1. 
%
In particular, unlike $F_f(T)$, $\Delta\phi_f(T)$ does not depend on the field-derivative
  of ${\cal A}$. 
Thus $\Delta\phi_f(T)$ is easier to interpret than $F_f(T)$: 
  when $\Delta\phi_f$ increases, the Fermi surface is expanding, and vice-versa. 

In Fig. \ref{fig-PhivsT}, we plot $\Delta\phi_f(T)$ 
  at several magnetic fields spanning the AFQ phase, 
  finding large, systematic and often non-monotonic temperature dependence. 
According to our intepretation of the $F_f(B)$ data in Fig.\ \ref{fig-FvsB}, 
  these changes reflect the 
relative occupancies
 of $\Gamma_1$ and \G4p. 

\section{Discussion} \label{sec:discussion}

It is important to emphasize that, although 
the
changes 
we observe in ${\cal A}(T,B)$ are 
  proportional to changes in $-M(T,B)$, the magnetization is not {\em causing} the changes in 
  ${\cal A}$.
What we see as a single quantum oscillation frequency is actually the superposition
  of oscillations from spin-up and spin-down branches of the Fermi surface,
  which have very nearly the
  same back-projected oscillation frequency \cite{shoenberg84}.
Changing the polarized magnetic moment on the localized $4f$ states will produce a 
  contribution to the 
  spin splitting of these branches of the Fermi surface, additional to 
  the Pauli paramagnetic spin splitting that the applied magnetic field alone would 
  produce. 
However, it will not change their 
  {\em average} back-projected dHvA frequency, which is what we observe. 
The field and temperature dependence of ${\cal A}(B,T)$ 
  must primarily reflect changes in the {\em charge} distribution on the Pr $f$-orbital as 
  the relative occupation of $\Gamma_1$ and $\Gamma_4^{(2)}$ changes. 
This is important because it means that, even if the crystal field levels 
  were both non-magnetic singlets, the Fermi volume measurement would still be sensitive to 
  changes in their occupancy, although 
the magnetisation
$M$ would not. 

If we examine
the temperature dependence of $\Delta\phi_f(T)$ in Fig.\ 
  \ref{fig-PhivsT},
much of this behaviour can be easily 
  understood within the existing model of the crystal field states 
  of PrOs$_4$Sb$_{12}$, 
  when we consider that the $\beta$-sheet expands when the 
  occupancy of $\Gamma_1$ increases at the expense of \G4p, and vice-versa. 
Thus, for magnetic fields below the lower boundary of the AFQ phase
  (at 4.25 T, for example) $\Delta\phi_f$ decreases monotonically (the 
  $\beta$ Fermi surface contracts) 
  with increasing $T$ because the \G4p state is becoming 
  thermally occupied at the expense of the $\Gamma_1$ ground state. 
For fields above the upper boundary of the AFQ phase (e.g.\ 13 T) 
  we see the 
  opposite behaviour because the relative positions of $\Gamma_1$ 
  and \G4p are now reversed: $\Delta\phi_f$ {\em increases} monotonically 
  (the $\beta$ Fermi surface expands) 
  with increasing $T$   
  because Pr $4f$ electrons are being thermally excited 
  from the \G4p ground state to the $\Gamma_1$ excited state. 

For magnetic fields corresponding to the AFQ region, 
  the behaviour of $\Delta\phi_f(T)$
  is more complicated, but we assume that it still reflects the 
  relative occupations of the crystal field states as we have
  described. 
For $4.7~T \lesssim B \lesssim 8.5~T$
  there is a maximum in $\Delta\phi_f(T)$ at   
  the AFQ thermal phase boundary $T_{AFQ}$.
This is clearly seen in Fig.\ \ref{fig:FvsT}, as well as 
in 
several curves in Fig.\ \ref{fig-PhivsT}. 
As noted in the introduction, 
  AFQ order involves a superposition at each 
  Pr site of the form $\sum_j a_{i,j}(B,T)\,|\Gamma_4^{(2),j}\rangle + b_i(B,T)\,|\Gamma_1\rangle $, 
  where $i=1,2$ indexes the AFQ sublattice and 
  $j=+,0,-$ indexes the triplet crystal-field sublevels (see Appendix 2). 
As the temperature increases in this field  and temperature range, the amplitude of the 
  AFQ order parameter decreases. 
%
This is 
  reflected in a decrease in the $a_{i,j}(B,T)$ and a corresponding increase in $b_i(B,T)$: 
  \G4p is higher in energy than $\Gamma_1$, so its incorporation 
  into the ground state costs crystal field energy that can only be compensated by 
  AFQ energy. 
Thus, as $T$ increases towards $T_{AFQ}$, the occupancy of $\Gamma_1$ increases at the 
  expense of $\Gamma_4^{(2)}$, 
  and the $\beta$ Fermi surface expands.  
For $T>T_{AFQ}$, AFQ order is 
  gone, and 
there is 
an incoherent, thermal superposition of the crystal 
  field states on the Pr sites.
%
The occupation of $\Gamma_1$ therefore {\em falls} 
  while that of \G4p rises, as 
  Pr 4f electrons are thermally excited into the \G4p state
  causing the $\beta$ Fermi surface to shrink and $\Delta\phi_f(T)$ to fall. 
The non-monotonic behaviour 
of  $\Delta\phi_f(T)$
thus 
reflects a change in regime from coherent superposition 
  to incoherent thermal occupation
of crystal field states. 

For $8.5~T \lesssim B \lesssim 11.6~T$, we would expect this situation
  to be reversed due to the crystal field level-crossing, 
  producing a {\em minimum} in  $\Delta\phi_f(T)$ at $T_{AFQ}$. 
  This minimum is observed up to 9.75~T, but above this field we
  see $\Delta\phi_f$, and therefore $\Gamma_1$ occupancy, monotonically increasing 
  with increasing $T$ at all temperatures, even within the AFQ phase. 
  We do not fully understand this
behaviour,
but it 
appears to 
signal a change in the AFQ ground state in this
  field region, perhaps to an admixture 
of the different $\Gamma_4^{(2)}$ 
  states with very little $\Gamma_1$, consistent with previous suggestions that the 
  ordered structure changes at a first-order
  phase transition within the AFQ phase (shown as the higher-field purple 
  line in Fig.~\ref{fig-phasedia})\cite{tayama03}.  

Temperature dependence of Fermi surface areas is a subject that has only rarely been 
  discussed with respect to quantum oscillations \cite{lonzarich74,yelland07,shoenberg84},
  and such a strongly temperature dependent phase, particularly a non-monotonic variation 
  of $\Delta\phi_f(T)$, has not to our knowledge been previously reported.  
This does not mean that temperature dependence of the Fermi surface is uncommon, 
  but the usual way of analysing dHvA oscillations using Fourier transforms 
  could cause such temperature dependent phases to have gone unnoticed. 
Application of Eq.\ \ref{eq-deltaphi} to the data in 
  Fig.~\ref{fig-PhivsT} illustrates how sensitive the dHvA phase is
  to variations in Fermi surface area:
a tiny change in ${\cal A}$ leads to a large change in $\phi_f$.
At 10~T, for example, $\Delta\phi_f$ reaches about 
  11 radians at 2.5 K, 
which corresponds
to a change in ${\cal A}$ of only 
  $\sim 2$\%. 

The high sensitivity of the dHvA technique, particularly at  
  millikelvin temperatures, has allowed us to observe some unexpected 
  behaviour in the $T\rightarrow 0$ K limit. 
Figure \ref{fig-PhivsT}b shows that there is a clear difference between 
  data outside the AFQ phase (at 4.25~T, 4.5~T, 12.5~T and 13.0~T), in which 
  $\Delta\phi_f(T)$ approaches $T = 0$ K with zero slope,  and 
data
  within the AFQ phase, which show a distinctly positive slope as 
  $T \rightarrow 0$ K. 
We have found similar behaviour with 
  the magnetic field parallel to the (100) direction (data
  not shown).
The sign of the 
$T \rightarrow 0$
slope is positive everywhere within the AFQ phase, 
  independent of whether 
the magnetic field is higher or lower than
  the singlet-triplet level crossing near 8 T. 
The 
change in the slope of $\Delta\phi_f(T)$ 
  at low temperature 
can be seen very clearly in Fig.~\ref{fig:FvsT},
setting in near 250 mK.

In a mean-field picture, the 
  AFQ order parameter should saturate as $T \rightarrow 0$ K, 
  so the naive expectation is that the occupancy of $\Gamma_1$ and 
  $\Gamma_4^{(2)}$ should stop changing with temperature.  
In terms of AFQ order, $\phi_f(T)$ should, therefore, 
  have a slope that goes to 
  zero
for $T \ll T_{AFQ}$,
rather than a slope that {\em increases}, as we observe.
Although the behaviour we observe is not expected, a
  hint as to its origin
  can be found in thermal expansion measurements, 
  which also 
%
showed
temperature dependence 
for  $T \ll T_{AFQ}$, a result that was
ascribed to 
  the nuclear hyperfine interaction \cite{oeschler04}.

The nuclear hyperfine interaction  
  normally has a negligible effect on crystal field levels, but is 
  enhanced in Pr compounds with singlet ground states \cite{mackintosh91}. 
Pr has one stable isotope, with a $J=5/2$ nucleus,
so
%
what we have been calling ``the ground state" is actually a 
  manifold of six hyperfine states, split primarily by the 
  hyperfine dipole interaction  $H_{HF} = A I\cdot J = A I_zJ_z + (I_+J_- + I_-J_+)/2$, 
  with $ A \sim 50$ mK \cite{aoki02b}. 
The $I_zJ_z$ term, and the fact that $J_z=0$ for $\Gamma_1$, means that 
  the hyperfine splitting will be proportional to the amount of 
  \G4p in the electronic ground state. 
A key point, however, is that the off-diagonal $I_+J_-$ and $I_-J_+$ terms 
  mix \G4p and $\Gamma_1$, 
  so that, within the hyperfine manifold, different states can have differing 
  weights of \G4p and $\Gamma_1$, with the hyperfine ground state expected to have 
  the most \G4p. 
Outside the AFQ phase this mixing is weak 
  because $\Gamma_1$ and \G4p are separated in energy, 
  so the perturbation is second order in $AI_+J_-$.  
Within the AFQ phase, however, the ground state is a superposition 
  of $\Gamma_1$ and $\Gamma_4^{(2)}$, so there is a first-order matrix element and the 
  mixing is much stronger, i.e.\ it is proportional to $A$, rather than $A^2$. 

Within this picture,  
the low temperature contraction of the $\beta$ Fermi surface 
  within the AFQ phase can be understood:  
  for temperatures above about 300 mK, all of the hyperfine states of the ground state 
  manifold are equally thermally populated, 
  but as $T$ falls towards 0 K the lower hyperfine states, which have more \G4p character, 
  become preferentially occupied.  
Because the $\beta$ Fermi surface `doesn't like' $\Gamma_4^{(2)}$, 
(as we have discussed above)
it shrinks.  
%
In Appendix 2 we present a toy model that illustrates and  
supports this interpretation.  

The observed effect on the Fermi surface 
  is only of the order of 0.08\% of ${\cal A}$, but is still easily observed  
  by quantum oscillations: 
At several 
magnetic
field values, the hyperfine interaction 
  is actually a stronger influence on the admixture of $\Gamma_1$ and \G4p than the AFQ order. 

We note that the hyperfine interaction 
  may have a profound effect on the phase transition line 
  at the quantum critical point marking the lower boundary of the AFQ phase, 
  since hyperfine mixing should stabilize the AFQ order. 
Thus we would expect that the phase line could show deviations from the 
  predictions of simple AFQ theory in the low mK temperature range. 
There may also be enhanced nuclear adiabatic demagnetization effects on 
  crossing the AFQ phase boundary. 

We are not aware of any other technique that can so sensitively detect 
  changes in electronic states in this low temperature regime, 
  although similar conclusions could perhaps be drawn from 
  similarly detailed thermal expansion measurements, if they were performed. 
We note that the exact energy splitting of the hyperfine levels will be very sensitive 
  to the electronic states that form the hidden order in a multipolar system, 
  so it may be possible, by detailed fitting of the temperature dependence, to test 
  models of hidden order electronic states. 
This is an interesting possibility in other hidden order systems.

\section{Conclusions} \label{sec:conclusions}

In conclusion, we have observed a qualitatively new 
effect in quantum oscillations, which arises 
from the temperature dependent expansion and 
contraction of a 
Fermi surface as the population of localized electronic states changes. 
This very sensitive method of measuring the temperature dependence of the 
Fermi volume 
  has allowed us to map the relative occupation of crystal field levels in 
PrOs$_4$Sb$_{12}$, including the change from Boltzmann-type thermal occupation 
to the coherent superposition of crystal field levels that signals 
AFQ order in this material.
It has also allowed us to observe new features of the hyperfine interaction
  in PrOs$_4$Sb$_{12}$, related to a jump in the hyperfine mixing of the 
  crystal field levels on entry into the AFQ phase.
This application of quantum oscillations may, in future, give 
  useful information about the nature of hidden order in other systems.

\vspace{1cm}

\noindent
{\bf Acknowledgements:}

We are grateful to John Sipe for helpful discussions. 
This research was funded by the National Science and Engineering Council of 
Canada, the Canadian Institue for Advanced Research, the Marie Curie Program of 
the European Science Foundation, and the 
U.S. Department of Energy, Office of Basic Energy Sciences, DE-FG02-99ER45748.\\

\begin{center}
{\bf Appendix 1: The effect of a field and temperature dependent Fermi volume on quantum oscillations}
\end{center}

If the extremal area ${\cal A}$ of a Fermi surface is field or temperature dependent,
  it will affect the quantum oscillations through the Onsager expression
  \begin{eqnarray}
    F = \frac{\hbar {\cal A}}{2\pi e},
  \end{eqnarray}
  where $F$ is the dHvA frequency appearing in the argument of the oscillatory
  term,  $\sin(2\pi F/B + \phi_\circ)$ 
(see Eq. \ref{eq:LK}).
Because $F$ appears in combination with $1/B$,
  it cannot be extracted from the observed quantum oscillation frequency \cite{vanruitenbeek82}.
To see this, consider a field sweep centered on a field $B_\circ$, and 
  expand ${\cal A}(B,T)$ to first order in $\delta B$:  
\begin{eqnarray}
\lefteqn{\hbar \frac{ {\cal A}(B,T) }{eB} + \phi_\circ \simeq } \nonumber \\
    & & \hbar \frac{ {\cal A}(B_\circ,T) +
                    (B - B_\circ)\partial_B{\cal A}(B,T)|_{B_\circ} }{eB} + \phi_\circ = 
            \\
    & &
         \hbar \frac{ {\cal A}(B_\circ,T) -  B_\circ\partial_B{\cal A}(B,T)|_{B_\circ} }{eB}
                +  \frac{\hbar}{e}\frac{\partial{\cal A}(B,T)}{\partial B}|_{B_\circ}  + \phi_\circ
                   \nonumber   \\
               &\equiv&
         2\pi\frac{ F_f(B_\circ,T)}{B} +
                    \phi'_\circ(B_\circ,T),
\end{eqnarray}
  where
  we have used the shorthand notation $\partial_B \equiv \partial/\partial B$. 
The frequency 
\begin{eqnarray} \label{eq:bproj}
  F_f(B_\circ)  \equiv \left(1 - B\frac{\partial}{\partial B}\right)\frac{\hbar {\cal A}}{2\pi e}
\end{eqnarray} 
  is called the ``back-projected" frequency, and it is the frequency that is observed 
  in a quantum oscillation measurement.  

Geometrically, the back-projected frequency is the intercept of the tangent of $F(B)$ 
  at $B=0$, as illustrated in the main panel of Fig.\ \ref{fig-FvsB}b. 
The blue curve gives a possible field dependence of the Fermi surface cross-sectional area: 
  we hypothesize that the $\beta$ Fermi surface shrinks continuously with increasing 
  field, but the slope is more negative within the AFQ region.  
Back-projection is shown for one point, at 4.4 T. 
The green dashed line is the tangent of $F(B)$ at that field, and its intercept at 
  $B=0$ gives $F_f$. 
In this model, the reason that $F_f$ jumps up when the AFQ phase is entered is  
  not because ${\cal A}$ itself changes suddenly, 
but rather 
because its slope changes suddenly, 
  causing the back-projection to jump.  
This hypothesis is supported by the field-dependence of the magnetization, which looks similar 
  to the blue curve (but with a positive slope) \cite{tayama03}.
  In particular, we note the striking correspondence between $F_f$ at 100 mK and 
  the red 
%
curve
below it in Fig.\ \ref{fig-FvsB}a, which is proportional to  
  $-(1-B\partial/\partial B)M(B)$. 
As noted in the text, 
  this correspondence strongly suggests that ${\cal A}(B) - {\cal A}(0)$ 
  is proportional to $-M(B)$. \\

\noindent
{\bf Temperature Dependent Fermi Volume} 

If an extremal area ${\cal A}$ of a Fermi surface is temperature dependent, the most useful analysis 
  of 
quantum oscillations is to examine the temperature dependence of the dHvA phase. 
It initially seems more promising to extract $F_f(T)$ 
  at each temperature by fitting
  \begin{eqnarray}
    A_f(T)\sin\left( 2\pi \frac{F_f(T)}{B} + \phi_f(T)
                  \right)    \label{eq:fitted0}
  \end{eqnarray}
  where $A_f(T)$, $F_f(T)$ and $\phi_f(T)$ are free parameters representing 
  the amplitude, frequency and phase of the oscillation.
However, 
in the case of PrOs$_4$Sb$_{12}$,
the rapid field dependence of the Fermi surface in some regions (e.g.\ at 
 the boundaries of the AFQ phase) led us to restrict our analysis 
  to very narrow field ranges, and 
fitting the frequency of only a few oscillations is inherently noisy,
  as can be seen in Fig.\ \ref{fig:FvsT}. 
Moreover, from Eq.\ \ref{eq:bproj} above, $F_f(T)$ gives a combination of the temperature dependence 
  of $\cal A$ and its derivative with
  respect to field, which is difficult, if not impossible, to deconvolve to arrive at 
  the temperature dependence of ${\cal A}$. 
If we instead fix $F_f$ at the value obtained by fitting the oscillations in the lowest-temperature 
  trace, 
  allowing only $A_f(T)$ and $\phi_f(T)$ to be free parameters at higher temperatures, 
i.e.\ fitting 
  \begin{eqnarray}
    A_f(T)\sin\left( 2\pi \frac{F_{f,\circ}}{B} + \phi_f(T)
                  \right)    \label{eq:fitted1}
  \end{eqnarray}
  to the $T>T_\circ$ data, where $F_{f,\circ} = F_f(B_\circ,T_\circ)$ and
  $T_\circ$ is base temperature,
  this produces much less noisy results (see Fig.\ \ref{fig:FvsT}). 
This approach has also been taken in some previous dHvA studies \cite{lonzarich74,yelland07}. 

The interpretation of the temperature dependent phase turns out to be surprisingly simple. 
Assume that the change in ${\cal A}$ at temperature $T$ relative to its value at base temperature,  
\(
    \Delta {\cal A}(T) \equiv
                  {\cal A}(T) - {\cal A}(T_\circ),
  \)
  is small compared with ${\cal A}(T)$ itself (in our case the ratio is
  less than about 2\%).
Then, showing the field dependence explicitly so that the back-projection 
  can be 
included
at the appropriate time, we write
\begin{eqnarray}
 \lefteqn{ \hbar\frac{ {\cal A}(B,T)}{eB} + \phi_\circ   = 
  \hbar\frac{ {\cal A}(B,T_\circ) + \Delta {\cal A}(B,T)}{eB} + \phi_\circ } \nonumber \\
     &\simeq&
        \hbar\frac{ {\cal A}(B,T_\circ)}{eB} +
            \hbar \frac{ \Delta {\cal A}(B_\circ,T)}{eB_\circ} + \phi_\circ \\
     &\simeq&
        2\pi \frac{  F_f(B_\circ,T_\circ)}{B} +
                   \hbar\frac{\Delta {\cal A}(B_\circ,T)}{eB_\circ}
                + \phi'(B_\circ,T_\circ).
\end{eqnarray}
That is, a small correction to the frequency of an oscillation can, over a range of a 
  few periods, be accurately be treated as 
  a phase shift\footnote{This can readily be verified by plotting together, for example, the 
  functions $\sin(2\pi F/B)$, $\sin(2\pi (F+\delta F)/B)$ and $\sin(2\pi F/B + 2\pi\delta F/B_\circ)$ 
  with e.g.\ $F = 1000$ T, $\delta F = 10$ T and $B_\circ = 6.05$ T, over the range $6.0 {\rm\ T} < B 
  < 6.1 {\rm\ T}$. }  
\begin{eqnarray}
\Delta\phi_f(B_\circ,T) =  \hbar\frac{\Delta {\cal A}(B_\circ,T)}{eB_\circ},
\end{eqnarray}
where $\Delta\phi_f(B_\circ,T) \equiv \phi_f(B_\circ,T) - \phi_f(B_\circ,T_\circ)$. 
So the temperature dependent term in the phase
  {\em directly} gives the temperature dependence 
  of the extremal area, with no contamination by the derivative.
Note, however, that due to back-projection we do not know the value of 
  ${\cal A}$ at $T_\circ$: 
  we know quite precisely by how much ${\cal A}$ {\em changes} with temperature, but we know 
  much less precisely the absolute value of ${\cal A}$, at a given field. 

\noindent
\begin{center}
{\bf Appendix 2: Interaction between hyperfine coupling and quadrupolar order}
\end{center}

Our dHvA results for PrOs$_4$Sb$_{12}$ clearly show a downturn in 
$\Delta\phi_f(T)$ as $T \rightarrow 0~K$ within the AFQ phase. For example, at $T \simeq 250$ mK     in Fig.\ \ref{fig:FvsT} (green triangles), and similarly 
in several curves in Fig.\ \ref{fig-PhivsT}.
The inset of Fig.\ \ref{fig-PhivsT} shows that the increased low-temperature slope 
  is confined to the AFQ phase. 
In this Appendix we describe a toy model of the coupling between the dipolar 
  hyperfine interaction and quadrupolar order that could explain this behaviour. 
The model is somewhat artificial because 
  it ignores the broadening of the crystal-field states by nearest-neighbour 
  interactions, and because we have put the AFQ order in by hand.  

\begin{figure}
\includegraphics[width=7.0cm]{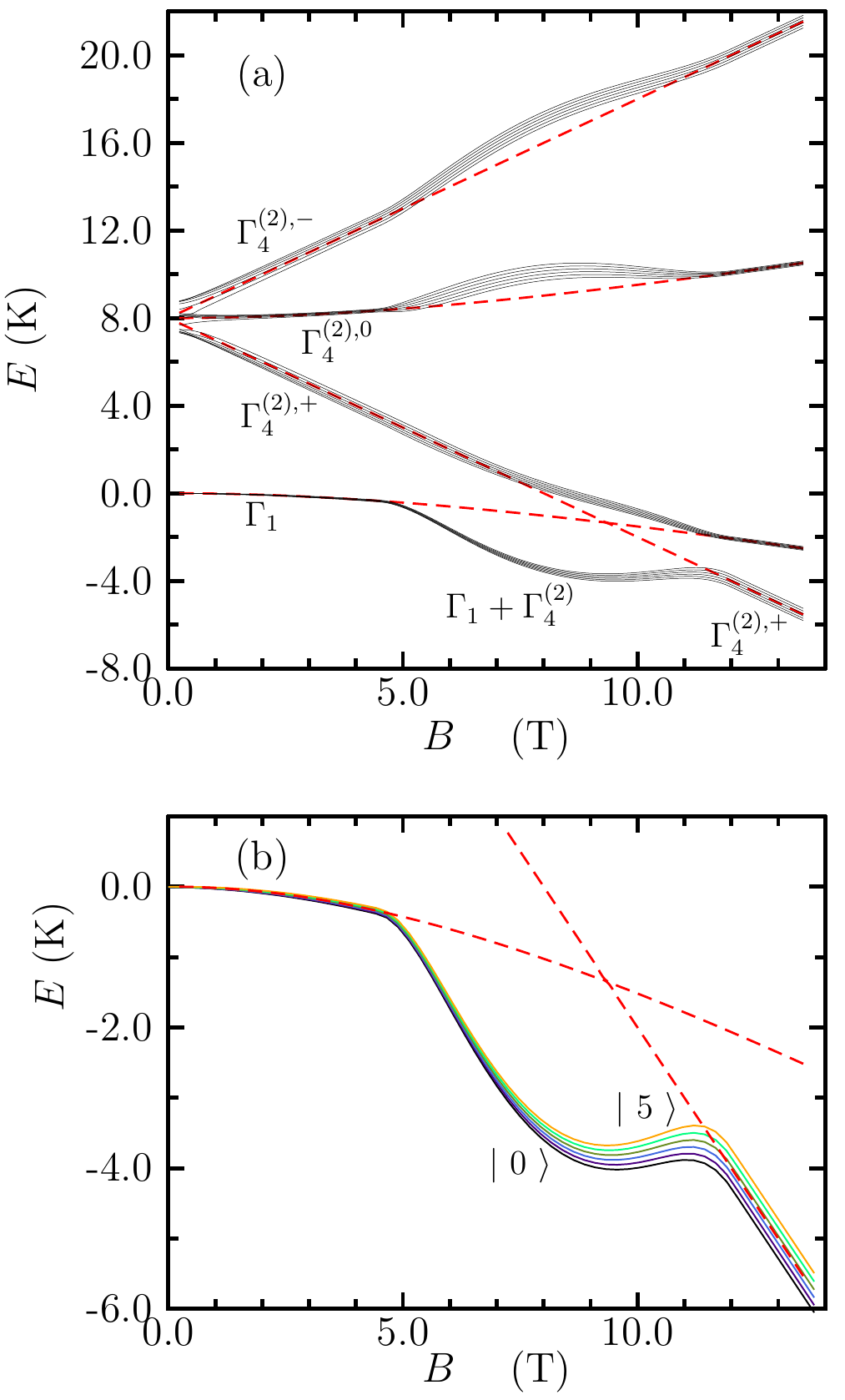}
\caption{ Energy levels and eigenstates for the two lowest crystal field levels 
of PrOs$_4$Sb$_{12}$ in the presence of a magnetic field, wth quadrupolar and hyperfine 
couplings.  (a) Energy vs.\ field. The red dashed lines are the 
  states of Eq.\ \ref{eq:xtalfield} with no quadrupolar order and the hyperfine interaction 
  turned off. 
The $-\delta h$ term mixes $\Gamma_4^{(2),0}$ with $\Gamma_1$ so its energy is  
  field dependent, despite $\Gamma_1$ being a singlet. 
Turning on quadrupolar order (Eq.\ \ref{eq:quadrupole})  
  between 4.75 and 11.5 T 
  avoids the level crossing of $\Gamma_4^{(2),+}$ and $\Gamma_1$, 
  and makes the ground state an admixture of $\Gamma_1$ and $\Gamma_4^{(2)}$ in this 
  field range.   
The hyperfine Hamiltonian (Eq.\ \ref{eq:hyperfine})  
  lifts the degeneracy of the six hyperfine eigenstates. 
(b) The ground state manifold of figure (a).  
The black line is the lowest hyperfine state, labelled $|0\rangle$, while the yellow line is the highest, 
  labelled $|5\rangle$. 
}
\label{fig-hyperfine}
\end{figure}

We use as our basis states ($\Gamma_1$, $\Gamma_4^{(2),+}$,   
  $\Gamma_4^{(2),0}$, $\Gamma_4^{(2),-}$). 
In this basis, the 
  crystal-field  Hamiltonian in the presence of a magnetic field 
  can be written \cite{shiina04} 
  \begin{eqnarray}
     {\cal H}_{cf} = \left( \begin{array}{cccc}
        0 & 0 & -\delta h & 0 \\
        0 & \Delta - h & 0 & 0 \\
        -\delta h & 0  & \Delta & 0 \\ 
        0 & 0 & 0 & \Delta + h 
                      \end{array}\right). \label{eq:xtalfield}
  \end{eqnarray}
We have used the notation of Shiina and Aoki \cite{shiina04b}, 
  where $h = g\mu_B\alpha H$ is the coupling of the $+$ and $-$ triplet 
  states to the applied field, $\delta = \beta/\alpha$ is a field-dependent 
  off-diagonal coupling between the singlet $\Gamma_1$ and the 
  $\Gamma_4^{(2),0}$ member of the triplet, and $\Delta$ is the 
  crystal field splitting between the singlet and the triplet at 
  zero field.  
The parameters $\alpha$ and $\beta$ are 
  $\alpha = 5/2 - 2 d^2$ and $\beta = 2\sqrt{5/3}d$, 
  where non-zero $d$ arises from the reduction of the symmetry of the Pr site from octahedral 
  to tetrahedral, and 
characterizes the resulting mixing of $\Gamma_5$ and 
  $\Gamma_4$ triplets that would be crystal field eigenstates in pure octahedral symmetry.
The eigenvalues of this Hamiltonian are shown by the dashed red lines in Fig.\ 
  \ref{fig-hyperfine}a. 
(This Hamiltonian may not be complete -- magnetoresistance 
  measurements on dilute  Pr$_{1-x}$La$_x$Os$_4$Sb$_{12}$ suggest that 
  the crystal-field level crossing is avoided even in the absence of 
  quadrupolar order \cite{rotundu07}.)  

To this we artificially add quadrupolar order between 
  4.75 and 11.5 T. 
We choose the so-called $X_y$ form of quadrupolar order 
  (see e.g.\ ref.\ \cite{kusunose09}): 
  \begin{eqnarray}
  {\cal H}_{Q} = \frac{1}{c1^2 + c2^2}\left( \begin{array}{cccc}
        0 & c1 & 0 & c1 \\
        c1 & 0 & c2 & 0 \\
        0 & c2  & 0 & -c2 \\ 
        c1 & 0 & -c2 & 0
                      \end{array}\right), \label{eq:quadrupole}
  \end{eqnarray}
  where we put in the AFQ order by hand by setting 
  \begin{eqnarray}
  c1 &=& \left\{ \begin{array}{ll} 
             c1_0(h - \frac{\Delta}{2.0})(\frac{3.5\Delta}{2.0} - h) & {\rm if}
                   \frac{\Delta}{2.0} < h < \frac{3.5\Delta}{2.0}   \\
             0             & {\rm otherwise}
                        \end{array}\right.  \label{eq:hyperfine}
  \end{eqnarray}
  where $c1_0 = 0.04\sqrt{35.0(1 - d^2)}$ and $c2 = 0.01\sqrt{3}(13-20d^2)c1/c1_0$. 
Between 4.75 and 11.5 T this term mixes the $\Gamma_1$ and the $\Gamma_4^{(2)}$ states, and 
  removes the $\Gamma_1/\Gamma_4^{(2),+}$ level crossing. 

When we introduce the hyperfine dipole interaction the Hamiltonian expands to a 
  24 $\times$ 24 matrix, with additional matrix elements of 
\begin{eqnarray*}
{\cal H}_{HF} = A\vec{I}\cdot\vec{J} = AI_zJ_z + \frac{A}{2}\left(I_{+}J_{-} + I_{-}J_{+}\right).
\end{eqnarray*}

\begin{figure}
\includegraphics[width=7.0cm]{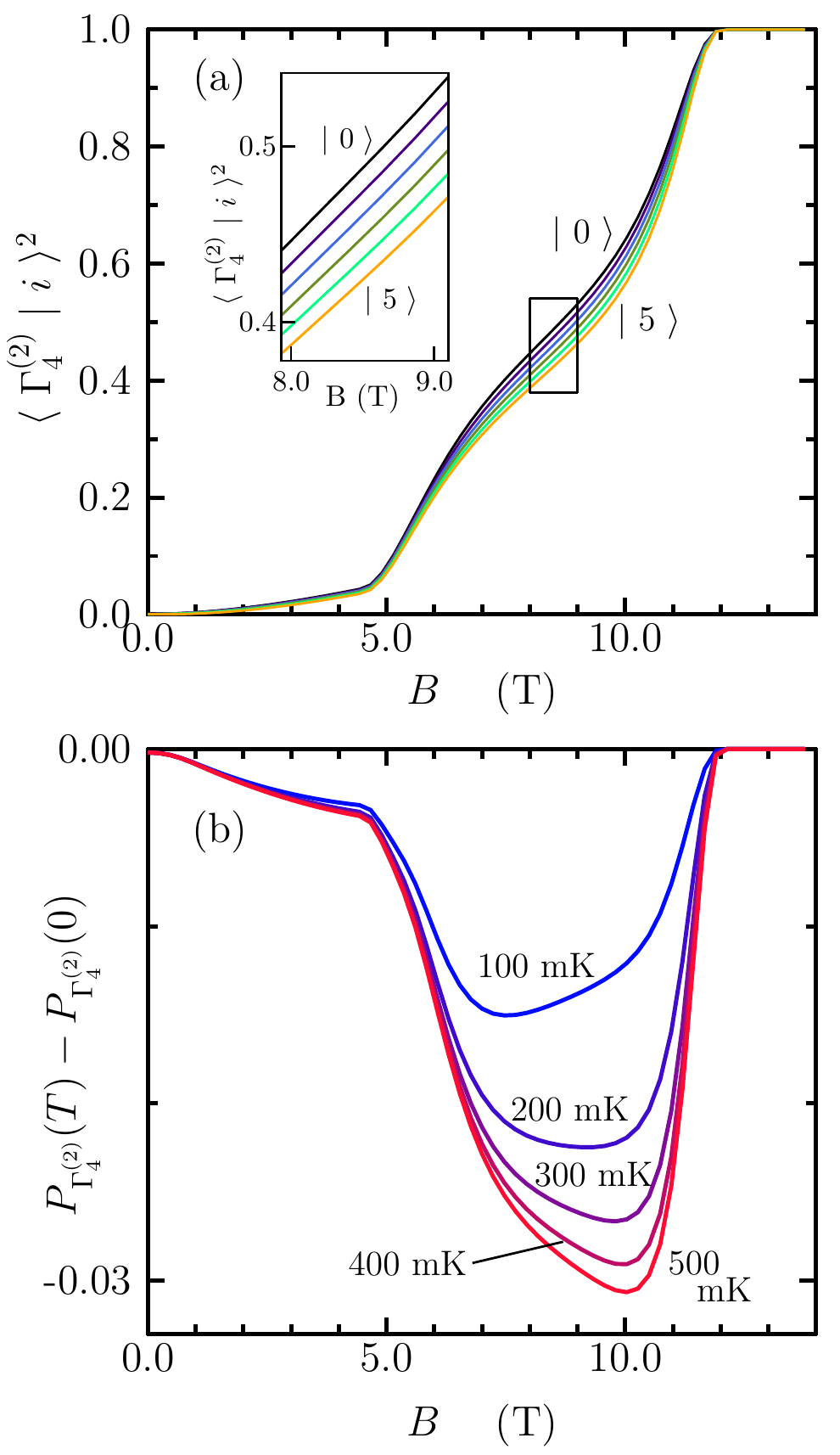}
\caption{ 
(a) The proportion of $\Gamma_4^{(2)}$ in the six hyperfine states of 
  the ground state manifold.  At $B=0$ all of the hyperfine levels in 
  the electronic ground state are purely $\Gamma_1$. 
At $B>11.75$ T, 
they are all purely $\Gamma_4^{(2)}$. 
In the AFQ region their composition is surprisingly different:  
  at a given field, the ground state, $|0\rangle$ has 
  the most $\Gamma_4^{(2)}$ character, and each successive excited hyperfine state 
  has progressively less.  The inset zooms in on the curves between 8 and 9 T.
(b) Change in thermal occupation of $\Gamma_4^{(2)}$ from its $T=0$ value, calculated by applying 
  a Boltzmann distribution to the eigenstates in (c). 
In the AFQ region, occupation of $\Gamma_4^{(2)}$ falls with increasing $T$ 
  because the excited hyperfine states, which have less $\Gamma_4^{(2)}$ character,  
  become thermally occupied. 
}
\label{fig-occupancy}
\end{figure}

Diagonalizing the full Hamiltonian (including crystal field, quadrupolar and hyperfine terms) 
  gives the black lines
in
Fig.~\ref{fig-hyperfine}a.  
Each of the electronic energy levels is split into six hyperfine levels. 
  Fig.\ \ref{fig-hyperfine}b focuses on the ground state manifold, and it can be seen 
  that the hyperfine splitting grows 
  rapidly through the AFQ region as the triplet $\Gamma_4^{(2),+}$ is progressively 
  mixed into the ground state manifold by the AFQ order.   
In effect, at a given magnetic field, the AFQ Hamiltonian 
  produces a certain admixture of $\Gamma_1$ and $\Gamma_4^{(2),+}$ in the ground state.
The more $\Gamma_4^{(2)}$ there is in the ground state, the larger the dipole term in the hyperfine 
  Hamiltonian, and thus the larger the splitting of the hyperfine states. 
However the hyperfine Hamiltonian, even though it is weak, 
  has a back-effect on the states through the $I_+J_-$ and $I_-J_+$ terms, 
  modifying 
  the admixture of $\Gamma_1$ and $\Gamma_4^{(2),+}$ 
  within the various hyperfine states of the ground-state manifold. 
Fig.\ \ref{fig-occupancy}a shows the result: the six hyperfine states in the ground-state 
  manifold have different amounts of $\Gamma_4^{(2)}$ character, 
  with the lowest state, $|0\rangle$, having up to about 5\% more $\Gamma_4^{(2)}$ at a given field 
  than the highest hyperfine state $|5\rangle$. 
As a result of these differing amounts of $\Gamma_4^{(2)}$, 
  when the temperature becomes low enough that the lower hyperfine states are preferentially 
  occupied, the occupancy of $\Gamma_4^{(2),+}$ increases, 
  causing the $\beta$ sheet of the Fermi surface to shrink.

In Fig.\ \ref{fig-occupancy}b, a Boltzmann average of the occupancy of $\Gamma_4^{(2)}$ 
  over the ground state manifold, at selected temperatures, allows us to plot 
  the {\em change} in $\Gamma_4^{(2)}$ occupancy relative to $T=0$~K as a function of $B$.
It can be seen that the change of $\Gamma_4^{(2)}$ occupancy with temperature is much stronger 
  within the AFQ phase.  
Moreover, it 
changes
in the same direction as 
suggested by
our data (decreasing weight of $\Gamma_4^{(2),}$ with 
  increasing temperature). 
The largest change occurs between 0 K and 100 mK, and the rate of change slows and becomes 
  quite small between 400 mK and 500 mK.  
This is quite similar to our observations, and leads us to believe that a more rigorous 
  model would provide good agreement. 
It also appears that 
a similar effect should be 
observed at temperatures well below 100~mK for fields below 4.75~T.
%
Above the AFQ phase, however, the ground state is 
purely $\Gamma_4^{(2),+}$, so there is no temperature dependent admixture in 
the ground state manifold in this region.

In this model, we have ignored the hyperfine quadrupole interaction, which is weaker than the 
  dipole interaction, but 
could 
also
produce an observable effect below 100 mK.  
Of course, 
the occupancies of the hyperfine states 
  will 
eventually 
saturate, and 
the Fermi surface will become temperature independent, but, 
  in PrOs$_4$Sb$_{12}$, this may not happen until low 
millikelvin 
temperatures are reached. 

Dependence of Fermi volumes on hyperfine levels 
is
an intriguing prospect 
  for future investigations of the physics of strongly correlated electron systems. 
From 
dHvA measurement such as we have described here,
it should be possible to extract detailed information about the coupling of 
  nuclear states to electronic energy levels, and thus determine the nature of the 
  electronic energy levels themselves. \\

\noindent
\begin{center}
{\bf Appendix 3: Comparison with thermal expansion measurements}
\end{center}

An obvious question regarding our results is whether the temperature dependence 
  that we have observed in the $\beta$ Fermi surface 
  might arise from simple changes in electron density 
  due to thermal expansion of the crystal. 

In free electron theory, the Fermi surface area is related to the 
  volume by  
  \begin{eqnarray}
    {\cal A} = \pi k_F^2 = \pi \left( \frac{3\pi^2 N}{V} \right)^{2/3}, 
  \end{eqnarray}
  where $k_F$ is the Fermi wave-vector, and $N/V$ is the conduction 
  electron density. 
Thus a change in sample volume of $\delta V$ will produce a corresponding 
  change of Fermi surface area of 
  \begin{eqnarray}
  \frac{\delta {\cal A}}{\cal A} = - \frac{2}{3} \frac{\delta V}{V}. 
    \label{eq:dkF}
  \end{eqnarray}
From data in reference \cite{oeschler04}, 
  at 6T with the field along (100) 
  the fractional volume change is 
  around $2.0\times 10^{-6}$ between 0.2 and 1 K. 
The dHvA phase changes by 
about 3 radians between 0 and 1 K 
under the same magnetic field conditions.
Using $\Delta A  = eB \Delta \phi/\hbar$, and the Onsager relation, 
  $F = \hbar {\cal A}/2\pi e$, we find that this translates to a 
  fractional change of the extremal area of $\sim +3\times 10^{-3}$, 
  which is three orders of magnitude 
  larger than the effect we would predict using equation \ref{eq:dkF}
and the data in reference \cite{oeschler04},
 and it has the wrong sign: 
  rather than expanding, the Fermi surface should shrink as 
  $V$ increases.  
  So the Fermi surface area change that 
  we observe 
does not arise 
from a simple change in electron density due to thermal 
  expansion of the crystal.


%

\end{document}